\documentclass{memsupp}
\usepackage{mynatbib}\usepackage{txfonts}\usepackage{balance}
\usepackage{graphicx}
\idline{22}{9}
\begin{document}
\def\Teff{$\rm T_{eff }$}
\def\kms{$\mathrm {km s}^{-1}$}
\def\Lise{$\rm^{7}Li$}
\def\Lisi{$\rm^{6}Li$}
\def\Feu {Fe\,{\small I}}
\def\Fed {Fe\,{\small II}}
\def\Halpha {$\rm H_{\alpha}$}

\title{
The cosmic lithium problem 
}

\subtitle{an observer's perspective}

\author{
M.\,Spite 
\and F.\,Spite
\and P.\,Bonifacio
}

\offprints{M. Spite}

\institute{
GEPI Observatoire de Paris, CNRS, Universit\'e Paris Diderot, F-92195
Meudon Cedex, France, 
\email{monique.spite@obspm.fr}}

\authorrunning{Spite}

\titlerunning{Lithium in 2012}

\abstract{
Using the cosmological constants derived from WMAP, the standard big bang nucleosynthesis (SBBN) predicts the light elements primordial abundances for $\rm^{4}He$, $\rm^{3}He$, D, \Lisi~ and \Lise. 
These predictions are in satisfactory agreement with the observations, except for lithium which displays in old warm dwarfs an abundance depleted by a factor of about 3. Depletions of this fragile element may be produced by several physical processes, in different stellar evolutionary phases, they will be briefly reviewed here, none of them seeming yet to reproduce the observed depletion pattern in a fully convincing way.

}
\maketitle{}

\section{Introduction}
 Over the past decade WMAP measurements of the cosmic microwave background (CMB) radiation allowed a precise determination of the cosmological constants:  \citet{LDH11}, \citet{KSD11}.  
As a consequence the cosmic baryon density of the Universe $\rm \Omega_{b}$ that, in the standard model governs the Big Bang nucleosynthesis (BBN) of the elements during the first few minutes in the evolution of the Universe, is now known with a good precision and therefore, the quantity of $\rm^{3}He,~ ^{4}He, D,~ ^{6}Li~ and~ ^{7}Li$ produced by the standard BBN.
 
These predictions may be compared with the primordial abundances of these elements inferred from observational data, but, at least for lithium ($\rm ^{7}Li~ and~  ^{6}Li$), they do not agree.
 
The aim of this meeting is to  explore the ways to reconcile predictions of the BBN to observations. In this introductory review we will try to provide an update of the ``lithium problem''.


\section{A Brief summary of the predictions of the standard BBN} \label{bbn}
Broad reviews of primordial nucleosynthesis and its relation to cosmology and lithium problems can be found e.g. in \citet{CFO08}, \citet{IMM09},  \citet{CocVan10}, \citet{Fields11}, \citet{CocGX12}, \citet{gary}.
 
In the standard BBN, the production of the elements depends only on the baryon to photon ratio $\eta$.
At low energy scales, the value of $\eta$ is simply related to $\rm \Omega_{b}$.
The recent seven-year WMAP data release leads to $\rm \Omega_{b} h^{2} = 2.249 \times 10^{-2}$ and $\eta = 6.19 \pm 0.15 \times 10^{-10}$ \citep{LDH11}.

With this value of $\eta $, BBN is expected to produce a quantity of lithium corresponding to 
$\rm A(^{7}Li) =  ~2.72$  
 \footnote{with the usual notation  A(Li)= $\rm \log (N(Li)/N(H)) +12$} ~and\\
$\rm A(^{6}Li) =  -2.00$ \\
with thus an extremely low  value of the isotopic ratio $\rm ^{6}Li/^{7}Li \approx 2 \times 10^{-5}$.

\section{How to observe the primitive Galactic matter ?}

In the first approximation, from the theory of stellar atmospheres, the chemical composition of the atmosphere of dwarf stars is a good witness of the chemical composition of the cloud that formed these stars. 
Therefore, if such low mass stars (with $M<0.9M_{\odot}$) were formed at the very beginning of the Galaxy, since their lifetime is larger than the Hubble time, they are still shining today and the chemical composition of their atmosphere is similar to the the chemical composition of the primitive cloud from which they have been formed.      

However, lithium is an exception because it is very fragile, destroyed as soon as the temperature is higher than $2.5 \times 10^{6}$ K for \Lise~  
and even $2.0 \times 10^{6}$ K for \Lisi~ through the reactions\\
$\rm^{7}Li +p \to  ~^{4}He +~^{4}He$\\
$\rm^{6}Li +D \to  \,^{4}He +~^{4}He$\\ 
$\rm^{6}Li +p \to  ~^{4}He +~^{3}He$\\

These temperatures are never reached in the stellar atmospheres but if the convection zone is deep, lithium in the atmosphere is swept along to hot deep layers and destroyed little by little;  its atmospheric abundance decreases. This happens for example in the Sun and in cool low mass stars. 

But in warm metal-poor stars, like turn-off stars, with $\rm T_{eff } \ge 5900\,K$ the convection is not so deep, and lithium should be preserved. Therefore, the lithium abundance in the atmosphere of these stars is expected to be a good witness of the abundance of lithium in the primitive Galactic matter. 

\begin{figure*}[t!]
\resizebox{\hsize}{!}{\includegraphics[clip=true]{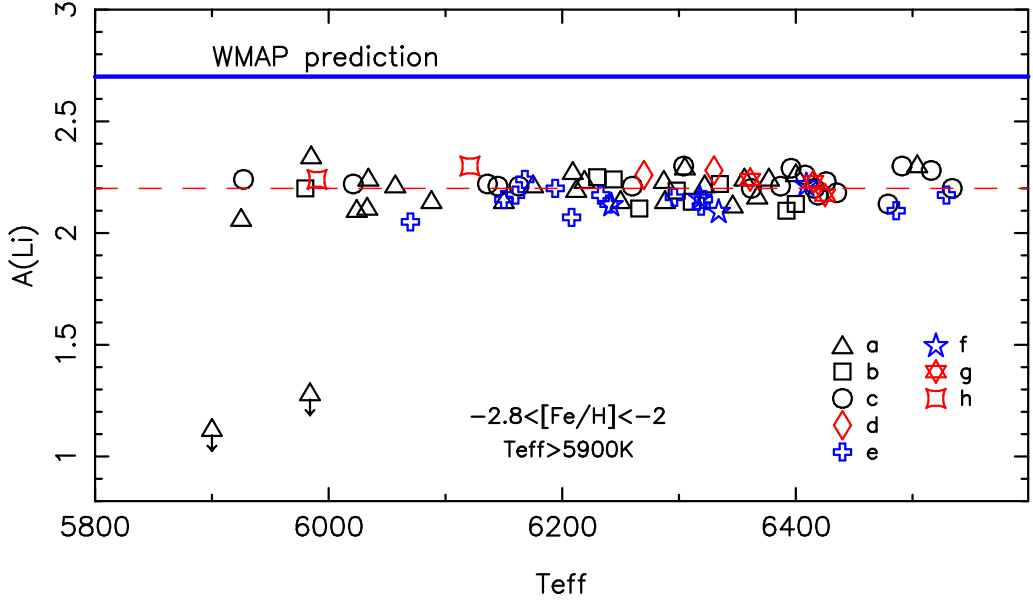}}
\caption{\footnotesize Lithium abundance vs. $\rm T_{eff}$ for stars selected for a temperature higher than 5900K (without deep convection) and without extreme deficiency $\rm -2.8<[Fe/H]<-2.0$.
Within these limits the abundance of lithium is independent of the temperature and the
metallicity: $\rm A(Li)\approx 2.2$ (lithium plateau). 
The symbols correspond to: a) \citet{ChaPr05}, ~~b) \citet{ALN06}, ~~c) \citet{MelCR10},  
~~d) \citet{AokBB09}, ~~e) \citet{HosRG09}, ~~f) \citet{BonMS07},  ~~g) \citet{SboBC10},
~~h) \citet{SchK12}. 
}
\label{plat}
\end{figure*}
%

\begin{figure*}[t!]
\resizebox{\hsize}{!}{\includegraphics[clip=true]{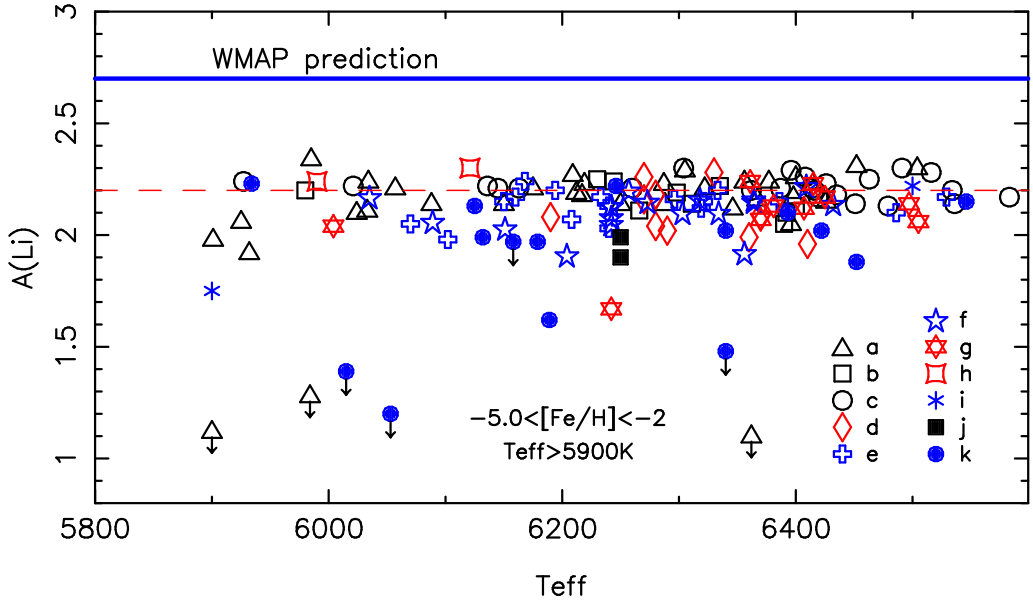}}
\caption{\footnotesize A(Li) vs. $\rm T_{eff}$ for stars with a temperature higher than 5900\,K (to avoid deep convection) and $\rm -5.0<[Fe/H]<-2.0$.
Many stars with $\rm [Fe/H]<-2.8$ are significantly below the lithium plateau, the scatter compared to Fig. \ref{plat} increases dramatically. 
The symbols are the same as in  Fig.\ref{plat} with in addition: ~~i) \citet{GonBL08}, ~~j) \citet{GarCR08}, ~~k) \citet{BonSC12}.  
}
\label{plat2}
\end{figure*}
%

\begin{figure*}[t!]
\resizebox{\hsize}{!}{\includegraphics[clip=true]{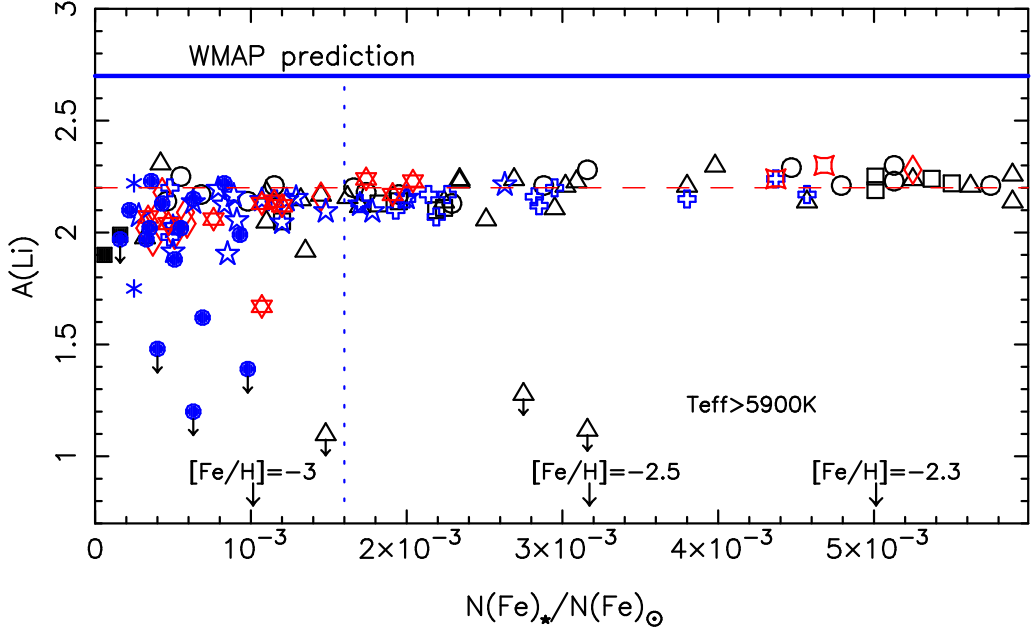}}
\caption{\footnotesize A(Li) vs. $\rm N(Fe)_{*}/N(Fe)_{\odot}$ 
(linear scale) for turn-off stars with a temperature higher than 5900\,K. 
For  $\rm N(Fe)_{*}/N(Fe)_{\odot}>1.6 \times 10^{-3}$ or  $\rm [Fe/H]>-2.8$ (dotted line), The abundance of lithium is almost constant but as soon as $\rm [Fe/H]<-2.8$ suddenly most of the stars are below the lithium plateau. 
The symbols are the same as in  Fig.\ref{plat} and \ref{plat2}.  
}
\label{LiFe}
\end{figure*}

\section {Abundance of \Lise~ in field Galactic stars}

Since lithium abundance in the atmosphere of metal-poor stars has been considered as one of the diagnostic to constrain the description of the primordial Universe, a wealth of data is available in the literature, and
in particular since 2005: \citet{ChaPr05}, \citet{ALN06}, \citet{AsMe08}, 
\citet{GonBL08}, \citet{GarCR08}, \citet{AokBB09}, \citet{HosRG09},
\citet{SboBC10}, \citet{MelCR10}, \citet{SchK12}.
The temperature is a crucial parameter to determine the abundance of lithium in the atmosphere of the stars. Sometimes the authors have used the colors of the stars, sometimes the dependance of the iron abundance on the excitation potential of the \Feu~ lines, and often the wings of the \Halpha~ line. \\

If we consider all the stars hotter than 5900\,K  and with a metallicity in the interval $\rm -2.8<[Fe/H]<-2.0$, the lithium abundance in this sample is constant: $\rm A(Li)\approx 2.2 \pm 0.06$ (Fig. \ref{plat}).  In spite of the inhomogeneity of the temperature determinations by various authors, the scatter is very small.\\ 
In these intervals of temperature and metallicity the lithium abundance does not depend on the metallicity nor on the temperature of the star. 
But this ``plateau'' is at a lithium abundance three times lower than the Big Bang prediction (see Fig. \ref{plat}).

In this range of temperature and metallicity,  two stars G\,122-69 and G\,139-08 are significantly below the lithium plateau. Both stars are well known lithium-poor stars \citep{Tho92}. \citet{Boes07} could measure the beryllium abundance in one of them (G\,139-08). The star is also Be-poor and thus is considered as a blue straggler.

Recently, with the advent of very efficient new stellar surveys (Hamburg-ESO, SDSS...) many extremely metal-poor (EMP) stars were discovered and thus the abundance of lithium could be measured in a number of turn-off stars with $\rm [Fe/H]<-2.8$.
If we add these EMP stars to our previous sample (Fig. \ref{plat2}),  many stars have a lithium abundance significantly lower than the "plateau", and the scatter compared to Fig. \ref{plat} increases dramatically, suggesting the influence of metallicity.

In figure \ref{LiFe} we have plotted the abundance of lithium in the same sample of stars as a function of $\rm ~N(Fe)/N(Fe)_{\odot}$ , with a linear scale. For more convenience we have added just above the X axis the corresponding values of [Fe/H]. The low scatter of the lithium abundance for $\rm N(Fe)/N(Fe)_{\odot} > 1.6 ~10^{-3}$ (or $\rm [Fe/H] > -2.8$, dotted line on the figure), is clearly seen with a mean value A(Li)=2.2. Then  for $\rm [Fe/H] < -2.8$ the plateau suddenly breaks down. But none of the stars has a lithium abundance significantly higher than the plateau.

If we want to solve the  \Lise~ problem
we have to explain that:\\
\\
$\bullet$ for $\rm -2.8 <[Fe/H]< -2.0$\\
the lithium abundance in turn-off stars with 
$\rm T_{eff } > 5900K$, 
is constant at a level
three times lower than the SBBN+WMAP prediction.\\
\\
$\bullet$ for $\rm [Fe/H]< -2.8$\\
the lithium abundance is strongly scattered 
below the lithium plateau. This scatter does  
not seem to depend on the temperature as  
soon as \Teff $ > 5900$\,K (Fig. \ref{plat2}).\\ 

\section {Is it possible to explain the behavior of \Lise~ by stellar processes ?}

\begin{figure*}[t!]
\resizebox{\hsize}{!}{\includegraphics[clip=true]{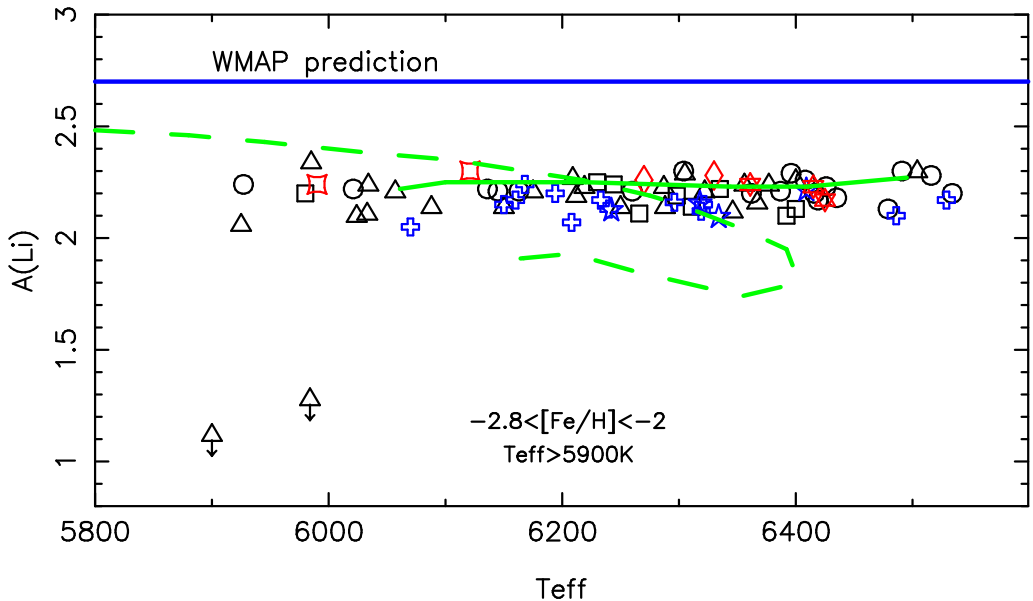}}
\caption{\footnotesize A(Li) vs. \Teff for stars with a temperature higher than 5900\,K and $\rm -2.8<[Fe/H]<-2.0$. 
The green dashed line represents the abundance of lithium in the atmosphere of stars with $\rm [Fe/H]= -2.3$ after 13.5Gyr.  The full line the abundance of lithium in presence of an ad hoc turbulence parameter.
}
\label{richard}
\end{figure*}

Two different ways were proposed to explain, in the frame of standard BBN, the unexpected behaviour of \Lise~ in metal-poor stars. Primordial Li would be depleted, either within the matter that formed the stars observed today, or during the lifetime of those stars.

\subsection{Destruction of Galactic \Lise~ in massive stars ?}
It was proposed by \citet{PiauBB06} that after a primordial production of lithium compatible with the WMAP predictions, the Galactic matter would have been rapidly processed in massive Pop~III stars in the mass range $10 M_{\odot}<M<40 M_{\odot}$ destroying two thirds of the primordial lithium.

Therefore, the EMP stars observed today would have been formed from a matter already depleted in lithium.\\
 But \citep{Pran07} remarked that these massive Pop~III stars would have produced heavy elements like carbon, oxygen... and thus the metallicity of the second generation of stars, (the old stars we observe today), would be much higher than the one observed in the EMP stars. Let us note that an ultra metal-poor star, without carbon enrichment has been found recently with [Fe/H]=-5.0 and thus with $Z/Z_{\odot} \approx 4.\times 10^{-5}$ 
 \citep{Caff11,Caff12}.\\
This interpretation that explains the avoidance zone (no lithium abundance in warm turn-off stars significantly higher than A(Li)=2.2, the value of the lithium plateau), is not able to explain the break of this plateau at a metallicity of $\rm [Fe/H]=-2.8$ in Fig. \ref{LiFe}.

On the other hand, the lithium abundance has been also measured in turn-off stars of $\omega$~Centauri by \citet{MonBS10}. This globular cluster is generally considered as the remnant of a dwarf spheroidal galaxy captured by the Milky Way, and thus, there is no reason that the same quantity of matter be processed by massive Pop III stars in $\omega$~Centauri and in the Milky Way. However the lithium abundance in the turn-off stars of $\omega$~Centauri is the same as in the field turn-off stars.\\
As a consequence, the interpretation of \citet{PiauBB06}, compatible with the Standard BBN,  does not seem to be well grounded.\\  

\subsection{Lithium plateau induced by turbulent diffusion ?}
Atomic diffusion is a slow gravitational  settling of the elements below the convective zone.
Following \citet{Mich84} \emph{"it is always present in stars. It cannot be turned off. It can only be rendered inefficient by sufficient mass motion either due to meridional circulation or turbulence"}. 
For a long time, it was supposed that diffusion was indeed rendered inefficient in metal-poor turn-off stars \citep{DelDK90}.

But \citet{RichMR05} computed the influence of diffusion on the lithium abundance with different turbulent parameters, after 13.5\,Gyr starting with a Li abundance compatible with the SBBN.\\ 
Without turbulence (Fig. \ref{richard}, dashed green line) the abundance of lithium would decrease when the temperature of the star increases in contradiction with the existence of a plateau. \\
But with an ad hoc model of turbulence (model ``T6.25'' where the turbulent diffusion coefficient, $D_{T}$ is 400 times larger than the He atomic diffusion coefficient at $\log T_{0} = 6.25$ and varies as $\rho^{-3}$) it is possible to represent the plateau (solid green line in Fig. \ref{richard}).  

Up to now, the cause of this turbulence is unknown. The diffusion coefficient is an ad hoc parameter, but at least with this parameter it becomes possible to represent the observed plateau with an original lithium abundance of 2.7 (as predicted by the standard BBN). Such a turbulence should exist to compensate the effect of diffusion. However, to date, turbulent diffusion does not explain that the plateau suddenly breaks down at $\rm [Fe/H] < -2.8$.

There were recently several attempts to determine this ``turbulent parameter'' in globular clusters: and in particular in NGC 6397: \citet{KorGR07}, \citet{GonBC09}, \citet{LinPC09}.
Since in globular clusters all the stars (at least the ``first generation'' stars) have the same age, the same metallicity and a well known evolutionary 
stage it is possible to check the small variations of the abundances as a function of the evolution of the stars. The result of these studies depends strongly on the adopted temperature scales, and up to now, they led to different turbulent parameters. In all these studies the estimation of the initial lithium abundance $\rm A(Li)_{0}$ is still not in complete agreement with the latest predictions of the SBBN. 

Moreover, if turbulent diffusion is responsible for the formation of a plateau of the lithium abundance at a level three times lower than the prediction of WMAP, we would expect that at least some stars (because, for example, they would have a little higher turbulence) would have kept a value of the lithium abundance close to the initial value and thus would lay in Fig. \ref{plat} and \ref{plat2}, between the WMAP value and the observed plateau. But no field star has been observed in this ``forbidden zone''. 

One turn-off star in NGC\,6397 \citep{KochLR11,Koch12}, and 
one in M\,4 \citep{MonVB12,Mon12} have been found Li-rich (respectively A(Li)=4.03 and A(Li)=2.87). But both stars are also Na-rich and are thus suspected to be polluted by the ejecta of a first generation of stars. The lithium abundance in these stars would not represent the abundance in the cloud which formed these globular clusters.

\begin{table*}
\caption{Comparison of the abundances of \Lise~and \Lisi~ (1D models) in \citet {ALN06} (VLT-UVES) or \citet {AsMe08} (Keck-HIRES)
and \citet{GarAI09} (Subaru-HDS).
}
\label{comp6li}
\begin{center}
\begin{tabular}{l@{~~~~~}c@{ }c@{ }c@{ }c@{ }c@{~~~~~}c@{ }c@{ }c@{ }c@{ }c}
\hline\hline
\\[-6pt]
             &\multicolumn{5}{c}{\citet{ALN06} or}                       & \multicolumn{5}{c}{\citet{GarAI09} }\\
             &\multicolumn{5}{c}{\citet{AsMe08}}     \\
Star         & \Teff    & $\log g$ & [Fe/H] & A(\Lise) & \Lisi/\Lise  & \Teff    & $\log g$ & [Fe/H] & A(\Lise) & \Lisi/\Lise\\
\hline
\\[-6pt]
BD +26 3578  & 6335& 4.0& $-2.26$ &2.22& $0.010\pm0.013$ & 6239& 3.9 & $-2.33$ & 2.21& $0.004\pm 0.028$\\
BD --04 3208 & 6298& 4.0& $-2.30$ &2.19& $0.048\pm0.019$ & 6338& 4.0 & $-2.28$ & 2.27& $0.047\pm 0.039$\\
G64-37       & 6368& 4.4& $-3.08$ &1.97& $0.111\pm0.032$ & 6318& 4.2 & $-3.12$ & 2.04& $0.006\pm 0.039$\\
\\[-6pt]
\hline
\end{tabular}
\end{center}
\end{table*}~~~~~

\section{Does the lithium abundance on the plateau represent the BBN production ?}

The observations could finally reflect a production of lithium by a non standard BBN at the level of the plateau.
Atomic diffusion would be inhibited within the temperature and metallicity range of the plateau (turbulence, meridional circulation. . .) as it was supposed, before the WMAP measurements \citep{DelDK90}. 

However even in this hypothesis, the breakdown of the plateau for  $\rm [Fe/H] < -2.8$ has to be explained.\\
We know that this depletion of lithium in stars with  $\rm [Fe/H] < -2.8$, is the result of an internal stellar depletion. \citet{GonBL08} indeed, could measure the abundance of lithium in the two components of CS~22876-32. This turn-off binary star is extremely metal-poor,  [Fe/H]=--3.6, and the abundance of observable elements (iron, calcium, oxygen etc.) have been found similar in both components. There is only one exception: the abundance of lithium is significantly different:
in  CS~22876-32A (\Teff=6500K) A(Li)=2.22, but in CS~22876-32B (\Teff=5900K) A(Li)=1.75 (see Fig. \ref{plat2} and \ref{LiFe}, where the two components of CS~22876-32 are represented by an asterisk).
A similar situation, albeit less extreme, has been found for the double-lined
spectroscopic binary BD~+26~2606 at [Fe/H]$=-2.5$ \citep{AokIT12,Aokmem12}.
The two components have temperatures of 6350\,K and 5830\,K and the lithium abundances are A(Li)=2.23 and 2.11, respectively. This may
suggest that the differential depletion between two stars of 
different mass may be present even at higher metallicities than that of 
CS~22876-32, but the  depletion is much less than that observed in CS~22876-32.
Alternatively this could simply mean that at metallicity --2.5 
the Li depletion by ``normal'' convection starts already around
5800\,K and for this reason the lower mass component has a slightly
depleted lithium content.

Since the components of binaries formed from the same cloud, the only possibility to explain this difference is that lithium has been depleted during the lifetime of the star (at least in CS22876-32B and BD~+26~2606B). A possible mechanism could perhaps be turbulent diffusion in the secondary, erased in the primary because of its evolution towards the subgiant phase. 

A number of possible non standard BBN have been proposed, producing a lower primordial lithium abundance, at the level of the plateau  
(see e.g.  \citealt{Oliv12} and \citealt{Kaj12} in 
this meeting). Such BBN would explain the plateau and 
also the strange ``forbidden zone'' described above. But any depletion along the plateau (by diffusion or other process) has then to be suppressed (by a strong turbulence\,? or adequate circulations such as convection, meridional circulation\,?  by adequate mixing\,?) At very low metallicities ($\rm [Fe/H] < -2.8$) aleatory depletions would be more or less at work (diffusion + turbulence\,? + others\,?)

\begin{figure*}
\resizebox{\hsize}{!}{\includegraphics[clip=true]{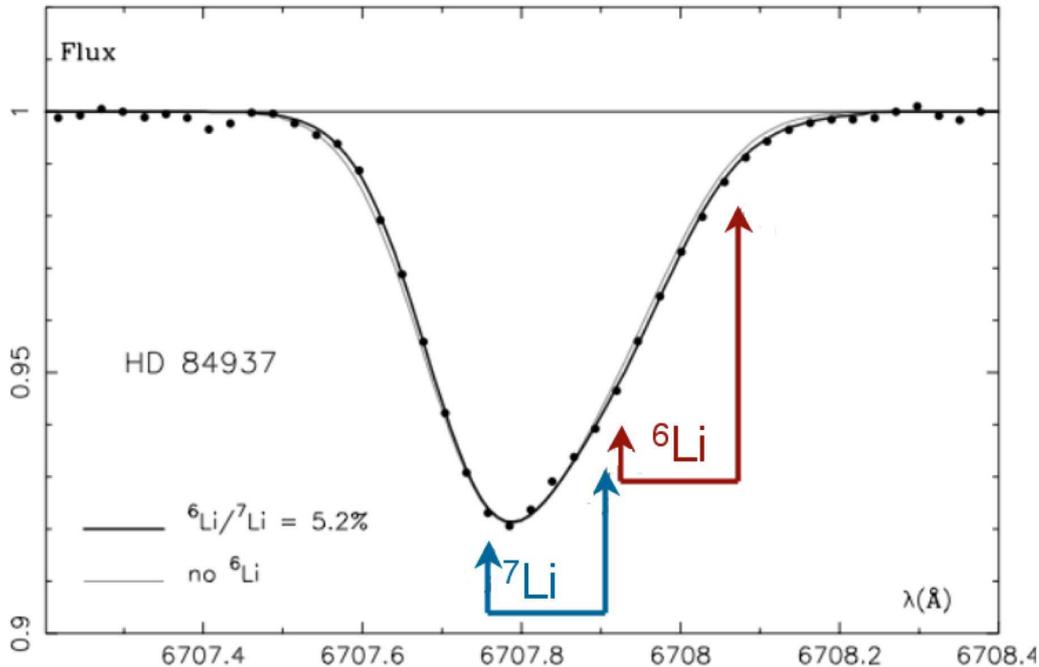}}
\caption{\footnotesize Profile of the lithium feature in the star HD 84937, \citet{Cay99}.
The dots represent the observed spectrum. The position of the \Lisi~and \Lise~ doublets are indicated. The thin grey line corresponds to an absence of \Lisi~ and the thick black line to a ratio 
\Lisi~/\Lise~ = 5.2\%.
}
\label{li6profile}
\end{figure*}

\section{Abundance of \Lisi~ in field metal-poor stars}

Several teams tried recently to measure the abundance of \Lisi~ in metal-poor halo stars, in particular:
\citet{ALN06}, \citet{StefCB09}, \citet{GarAI09}, \citet{Stef12}. This is a very difficult exercise because the lines of \Lisi~ and \Lise~ are superimposed and that, moreover, the hyperfine structure of the lines of \Lise~ (widening the profile) has to be taken into account (Fig. \ref{li6profile}). As a consequence very high quality spectra are needed: a resolution $R \approx 100\,000$ with a  $S/N \approx 500$ are necessary. 

\begin{figure}[]
\resizebox{\hsize}{!}{\includegraphics[clip=true]{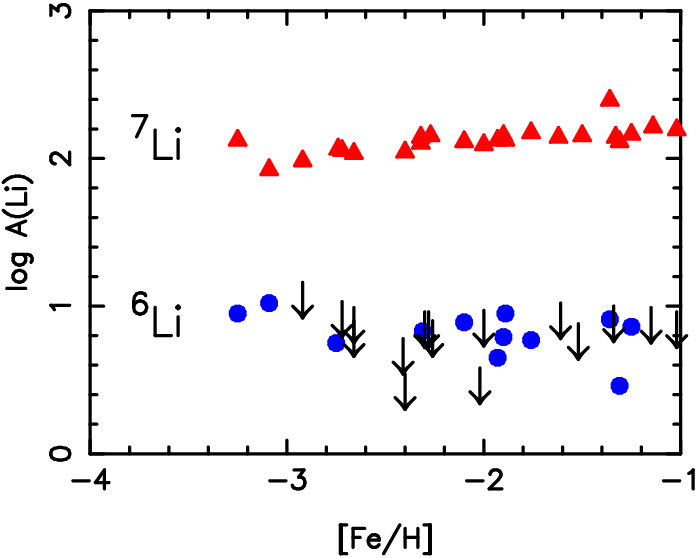}}
\caption{\footnotesize 
abundance of \Lise~ (red triangles) and \Lisi~ (blue filled circles) in a sample of 27 metal-poor stars observed by \citet{ALN06} and \citet{AsMe08}.
}
\label{aspli6}
\end{figure}

\citet {ALN06} and \citet {AsMe08} measured the lithium abundance in a sample of 27 metal-poor stars and they claimed that they have detected \Lisi~ in twelve of them (Fig. \ref{aspli6}). In all the stars the abundance of \Lisi~ was compatible with A(\Lisi)= 0.8, a value corresponding to a ratio \Lisi~/\Lise~ = 5\%, not compatible with the predictions of the standard BBN ($\rm ^{6}Li/^{7}Li \le 2 \times 10^{-5}$,  see section \ref{bbn}).
This \Lisi~ could have been produced by cosmic rays (spallation) after the BBN. But in this case we would expect that the abundance of \Lisi~ increases with the metallicity as it is for Be or B formed also by spallation. The constancy of the abundance of \Lisi~ is not compatible with a spallation origin of the \Lisi~observed in metal-poor stars.

\citet{CaySC07}, \citet{StefCB09} and \citet{Stef12} have shown that, when lines asymetries generated by convection are taken into account, the resulting abundance of \Lisi~ is strongly reduced.
 
On the other hand, \citet{GarAI09} have measured with the HDS spectrograph mounted on the Subaru telescope ($R\approx95000$, $S/N=500$) the \Lise~ and \Lisi~ abundances in five turn-off stars. In all these stars the abundance of \Lisi~ is compatible with an absence of \Lisi.  Three stars are in common with \citet {ALN06} and \citet {AsMe08}: BD\,+26 3578 (HD\,338529), BD\,--04 3208 (G\,13-09) and G\,64-37.
These observations are compared in Table \ref{comp6li}. In particular G\,64-37, was found  \Lisi-rich  by \citet {AsMe08} with \Lisi~/\Lise~ = 0.11 $\pm 0.03$, when \citet{GarAI09} found for this star \Lisi~/\Lise~ = 0.01 $\pm 0.04$\%: no \Lisi.
  
\citet{GarAI09} conclude that the abundance of \Lisi~ is very sensitive to the assumptions made for the continuum location, the residual fringing treatment and even the wavelength range covered in the analysis, and that the observational error is often underestimated.

At this conference \citet{Lind12} have shown some new measurements, based
on high quality spectra. Their analysis
method relies on lines other than \ion{Li}{i}
to estimate the rotational broadening, when such lines are treated in 
NLTE, the resulting broadening implies that no \Lisi\ is present
in the stars analysed by them. 

To date, the presence of \Lisi~ in the most metal-poor turn-off stars of our Galaxy does not seem to be firmly established.

\balance
\section {Conclusion}
The considerable observational progress (large telescopes, high resolution spectra, multiplex spectrographs, large surveys of metal-poor stars detecting many extremely metal-poor starsÉ) have specified the contour of the lithium problem, but no completely satisfactory solution has been found yet. Additional observations will be useful. The detailed observations of binary stars will certainly bring essential data. Precise distances (GAIA) and therefore luminosities, evolutionary status and ages of field stars will bring essential complements to the available information about the lithium problem.
 
\begin{acknowledgements}
It is a pleasure to  acknowledge
many helpful discussions with 
R. Cayrel about Li abundances.
\end{acknowledgements}

\bibliographystyle{aa}

\end{document}